\documentstyle[twoside,fleqn,espcrc2]{article}
\newcommand{\simm}[1]{\stackrel{#1}{\scriptstyle\sim}}
\title{ Lattice QCD Calculation of Hadron Scattering Lengths
    \thanks{presented by Y.~Kuramashi}}
\author{
Y.~Kuramashi
\address{
      National Laboratory for High Energy Physics(KEK), Ibaraki 305, Japan},
M.~Fukugita
\address{
      Yukawa Institute for Theoretical Physics, Kyoto University, Kyoto
 606, Japan},
H.~Mino
\address{
      Faculty of Engineering, Yamanashi University, Kofu 400, Japan},
M.~Okawa$^{\rm a}$
and
A.~Ukawa
\address{
      Institute of Physics, University of Tsukuba, Ibaraki 305, Japan},  }
\begin{document}

\begin{abstract}
Method of  calculating hadron multi-point functions and
disconnected quark loop contributions which are not readily
accessible through conventional techniques is proposed.  Results are
reported for $\pi$-$\pi$, $\pi$-$N$ and $N$-$N$ scattering
lengths and the flavor singlet-non singlet meson mass splitting
estimated in quenched QCD.
\end{abstract}
\maketitle
%
\section{Introduction}

Numerical simulation of lattice QCD has been applied to an
increasingly larger variety of strong interaction observables over the
years.  Yet these quantities all share the feature that their
calculation can be reduced to that of connected hadron 2-point
functions.  A number of physically interesting quantities do not fall
into this class: scattering amplitudes
requiring hadron 4-point functions are a prime example.
Amplitudes involving disconnected quark loops such as the flavor singlet meson
propagator and $\pi$-$N$ sigma term represent another important example.
Technically the difficulty stems from the fact that calculation of these
amplitudes requires quark propagators connecting arbitrary pairs of space-time
sites.  With the conventional method of point source the necessary number of
quark matrix inversions equals the space-time lattice volume, which would
require a prohibitively large amount of computer time.

We found that the
wall source technique\cite{Kenway,Kilcup}, in particular  {\em without
gauge fixing} as were employed in the original proposals of extended
sources\cite{Kenway}, could be effectively used to overcome the problem with a
modest cost of computing power\cite{pipi}.  The method has been applied to
calculate the full $\pi$-$\pi$, $\pi$-$N$ and $N$-$N$ 4-point functions at
vanishing relative momentum, from which we extracted the $s$-wave scattering
lengths with the help of  the relation between the two-particle energy in a
finite box and the scattering length\cite{HMPR,Luscher}.  With a slight
extension the method also allows an efficient  calculation of the two quark
loop
contribution to the flavor singlet $\eta'$ propagator.

In this report we present the results
together with some details of the technique.  All of the simulations
have been made within  quenched QCD at $\beta=5.7$  mostly  employing the
lattice size of $12^3\times 20$.
%
\section{Calculational technique}

Consider the box diagram contributing to $\pi$-$\pi$ scattering with
zero-momentum projected pion operators placed at the four time slices $t_i
(i=1,\cdots, 4)$.  To evaluate this amplitude on an $L^3\times T$ lattice, we
calculate $T$ quark propagators $G_t(n)$ with a wall source placed at the
time slice $t=1,\cdots, T$.  With these propagators we form
\begin{eqnarray}
\lefteqn{\quad
\sum_{\vec{x}_2,\vec{x}_3}\langle {\rm Tr}\big(
G^{\dagger}_{t_1}(\vec{x}_2,t_2)G_{t_4}(\vec{x}_2,t_2)
} \nonumber \\
\lefteqn{\qquad\qquad
G^{\dagger}_{t_4}(\vec{x}_3,t_3)G_{t_1}(\vec{x}_3,t_3)\big)\rangle , }
\label{eq:crec}
\end{eqnarray}
which equals the box amplitude except that the pion operators at the time
slices
$t_1$ and $t_4$ have to be taken to be non-local in space without insertion of
gauge link factors.

Fixing gauge configurations to some gauge is usually employed to
deal with the non-locality\cite{Kilcup}.  A problem with this method is
that excited states such as $\rho$ may contaminate signals for small time
intervals.  Alternatively one may take the gauge field average {\em without
gauge fixing} since gauge variant noise should
cancel out in the average.  One might worry that the noise overwhelms the
signal in practice since for each time slice $t_1$ and $t_4$ there are
$O((L^3)^2)$ gauge dependent non-local terms relative to $O(L^3)$ local gauge
invariant ones.  It is our experience, however, that the noise level is
controllably low at least for the pion.

Calculating $T$ quark propagators has another advantage.  For two hadron
operators placed at the same time slice, as is necessary if only the quark
propagator for a wall source at $t=0$ is available,
color Fierz rearrangement of quark lines takes place.  This leads to a mixing
among amplitudes having different quark line topologies, which is numerically
not straightforward to disentangle\cite{SGKP}.  Such a complication can be
trivially avoided with the present method by placing hadron operators at
different time slices.

The present  method can apply not only to a calculation of 4-point functions,
but
also to more general class of evaluation of $n$-point functions of hadrons.
%
%
\section{$\pi$-$\pi$ scattering lengths}

We have applied our method to calculate the $I=0$ and 2 $\pi$-$\pi$ scattering
lengths for both  Kogut-Susskind and Wilson quark actions, of which the
Kogut-Susskind result has already been published\cite{pipi}.
The procedure leading from $\pi$-$\pi$ 4-point
functions to scattering lengths is as follows.  The energy shift of
the two-pion state $\delta E=E-2m_{\pi}$ is extracted from the ratio
of 4- to 2-point
functions,
\begin{eqnarray}
R(t)&=&\frac{<\pi(t+1)\pi(t)\pi(1)\pi(0)>}
{<\pi(t)\pi(0)><\pi(t+1)\pi(1)>}\nonumber\\
 &=&Z\exp(-\delta E t).
\end{eqnarray}
In practice we use a linear fit $Z (1-\delta E t)$ in our quenched simulation
since $O(t^2)$ terms are generally not correct in the absence of dynamical
quark
loops.  The fitted values can then be converted to $s$-wave scattering length
$a_0$ using the
L\"uscher's relation\cite{Luscher}:
\begin{equation}
E-2m_{\pi}=-\frac{4\pi a_0}{m_{\pi}L^{3}}(1+c_{1}\frac{a_0}{L}
+c_{2}(\frac{a_0}{L})^{2})+O(L^{-6}) \label{eq:Lusc}
\end{equation}
with $c_{1}=-2.837297,c_{2}=6.375183$.

The two
calculations respectively used 160 ($m_qa=0.01$;KS) and 70 ($K=0.164$;Wilson)
$12^3\times 20$ configurations {\it without gauge fixing}. The $I=2$ amplitude
has been studied in pioneering work\cite{GMP,SGKP}. The calculation for
$I=0$ is more difficult since box and double annihilation diagrams with
quarkless intermediate state, both requiring a full application of our method,
contribute.  We found that
the  double annihilation amplitude is very small, consistent with chiral
perturbation theory and the OZI rule, and that the $I=0$ amplitude exhibits a
clear signal for attraction.

\begin{figure}[t]
\input psfig
\centering{
\hskip -0.0cm
\psfig{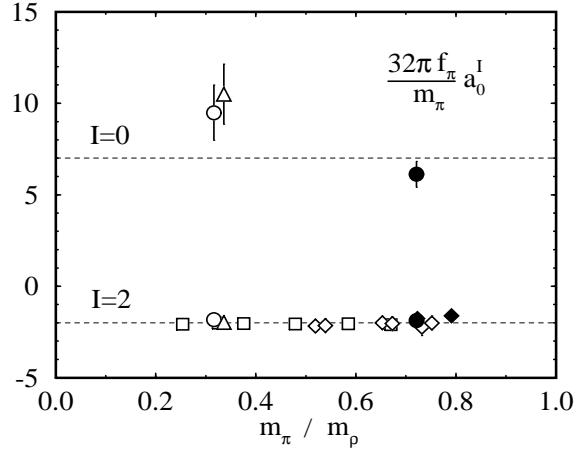}
\vskip -1.0cm  }
\caption{$I=0$ and 2 $s$-wave $\pi$-$\pi$ scattering lengths $a_0^I$.  Filled
and
open symbols denote Wilson and Kogut-Susskind results. Triangles are for
Coulomb gauge fixing.
Squares ($\beta=5.7)$ and diamonds ($\beta=6.0$) for $I=2$ are from
Ref.~[6].  Dotted lines indicate predictions of current algebra.}
\label{fig:pipisummary}
\vspace{-5mm}
\end{figure}

Our results for the scattering lengths are summarized in
Fig.~\ref{fig:pipisummary} together with those of Ref.~\cite{SGKP} for $I=2$.
Triangles are the results obtained with Coulomb gauge fixing described below.
We use $m_\pi$ and $f_\pi$  measured on the same set of configurations,
corrected by the improved $Z$ factor\cite{LM} for $f_\pi$ for the Wilson
case ($Z=1$ for the KS case since the conserved current was employed).
We observe an agreement of lattice results with current algebra within 1--2
standard deviations for both $I=0$ and 2 channels.  It is somewhat
unexpected that the agreement persists up to a quite heavy quark with
$m_\pi/m_\rho\approx 0.7-0.8$.

The errors shown in Fig.~\ref{fig:pipisummary} are statistical only.
For the Kogut-Susskind action the lack of
degeneracy of pions in the  Nambu-Goldstone and other channels invalidates the
$O(L^{-5})$  term in (\ref{eq:Lusc})\cite{SGKP}. This leads to an uncertainty
of 10\% in the $I=0$ result ( that for the $I=2$ result is less than 1\%).
Another source of
systematic error is violation of scaling due to a fairly large  lattice
spacing
of our simulation ($a\approx 0.2$ fm at $\beta=5.7$ ).  The results of
Ref.~\cite{SGKP} obtained at $\beta=5.7$ and 6.0 suggest, however,  that this
effect may be small, at least for $I=2$.

We have repeated the calculation with Coulomb gauge fixing employing 60 gauge
configurations, and found the results for $R(t)$ to be completely  consistent
with those obtained without gauge fixing on the same set of configurations.
Comparing the two results, we found that the magnitude of errors is
larger for the non-gauge fixed case.  The amount of
increase of errors, however, is
contained at the level of a factor of 1.5--2 times those for the Coulomb gauge
fixing, showing that gauge variant noise does not give rise to a serious
problem
for calculation of  $\pi$-$\pi$ 4-point functions.
The results for scattering length obtained with Coulomb gauge fixing are
plotted in Fig.~\ref{fig:pipisummary}.

All the results in Fig.~\ref{fig:pipisummary} are obtained in
quenched simulations, and hence do not include effects of dynamical quarks.
A potential problem with the quenched calculation is that it could be
affected by infrared singularities due to $\eta'$ loops suggested
recently\cite{BGS}.  Since the appearance of such singularities is a general
issue in quenched QCD, we shall discuss it in the subsequent section, in
connection with the $U(1)$ problem.  The problem of this unphysical
singularity anyway does not arise if calculations are made on full QCD gauge
configurations\cite{fimou}, which is a straightforward task to be carried
out.
\vspace{-3mm}

\section{$\eta'$ meson mass in quenched QCD }

The problem of a large mass splitting between the flavor singlet $\eta'$ and
the pion octet is known as the $U(1)$ problem.  In the large $N_c$ expansion
the
mass splitting $m_0^2=m_{\eta'}^2-m_\pi^2$ arises from iteration of virtual
quark
loops in the $\eta'$ propagator, each loop giving a factor
$m_0^2/(p^2+m_\pi^2)$.  In quenched QCD the series terminates at two
quark loops.  This implies that the magnitude
of $m_0$ may be estimated through a comparison of the two quark loop
amplitude having a double pole $m_0^2/(p^2+m_{\pi}^2)^2$ with the
one quark loop amplitude with a single pole $1/(p^2+m_{\pi}^2)$; the
ratio of the two amplitudes, each projected onto zero spatial momentum, is
expected to behave as
\begin{equation}
R(t)=\frac{<\eta'(t)\eta'(0)>_{\mbox{2-loop}}}
{<\eta'(t)\eta'(0)>_{\mbox{1-loop}}}\approx\frac{m_0^2}{2m_\pi}t.
\label{eq:etapropagator}
\end{equation}

It has been suggested that the double pole in the $\eta'$
propagator with a mass degenerate with that of $\pi$ leads to
infrared divergences which are not suppressed by powers of the pion mass
toward
the chiral limit in quenched QCD\cite{BGS}.  To one-loop order in chiral
perturbation theory, for instance, the pion mass receives a correction of
the form,
\begin{equation}
(m_{\pi}^{{\rm 1-loop}})^2=m_\pi^2\left(1-\frac{m_0^2}{8N_{f}\pi^2 f_{\pi}^2}
{\rm ln}\frac{m_{\pi}}{\Lambda}\right).
\label{eq:chirallog}
\end{equation}
The problem for $\pi$-$\pi$ scattering could be even more
serious\cite{BGSU}.  The double annihilation diagram with two quark loops can
be deformed into an $\eta'$ loop diagram with two insertions of the $m_0^2$
vertex.  Calculating the diagram in chiral perturbation theory, one finds
a divergent imaginary part at threshold in the $s$ channel, and in the $t$
channel a contribution of the form,
\begin{equation}
\delta E=\frac{1}{384\pi^2}\frac{m_0^4}{N_f^2m_\pi^2f_\pi^4}\frac{1}{L^3}
\label{eq:pipihairpin} \end{equation}
to the energy shift of the two pion system, which diverges quadratically as
$m_\pi\rightarrow 0$.
Whether such terms affect results of quenched simulations depends on
the magnitude of $m_0$.   A direct quenched estimate of $m_0$
through (\ref{eq:etapropagator}) is also important in this respect.

The two quark loop amplitude needed in (\ref{eq:etapropagator}) can be
evaluated
by the technique discussed above.   We solve for the quark propagator
with unit source {\em at every space-time site  without gauge fixing}. (Note
that this is crucial for the present case.)  With this quark propagator
$G(\vec{n},t)\equiv \sum_{(\vec{n}'',t'')}G(\vec{n},t;\vec{n}'',t'')$ we form
the expression,
\begin{equation}
\sum_{\vec{n},\vec{n}'}{\rm Tr}\{G(\vec{n},0)\gamma_5\}
{\rm Tr}\{G^{\dagger}(\vec{n}',t)\gamma_5\}.
\label{eq:2loop}
\end{equation}
This equals the two quark amplitude projected onto zero spatial momentum up to
gauge-variant non-local terms which, however,  cancel out in the ensemble
average.  A very nice feature of this technique is that it requires only a
single
quark matrix inversion for each gauge configuration.

\begin{figure}[t]
\input psfig
\centering{
\hskip -0.0cm
\psfig{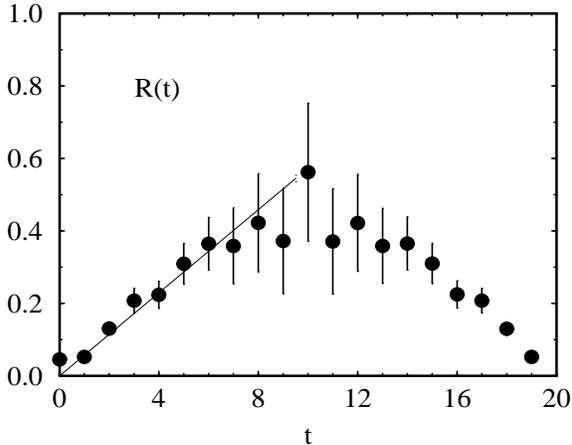}
\vskip -1.0cm  }
\caption{Ratio of two- and one-quark loop contribution to the $\eta'$
propagator
at $K=0.1665$ and $\beta=5.7$ on a $12^3\times20$ lattice.}
\label{fig:etaratio}
\vspace{-3mm}
\end{figure}

\begin{figure}[t]
\input psfig
\centering{
\hskip -0.0cm
\psfig{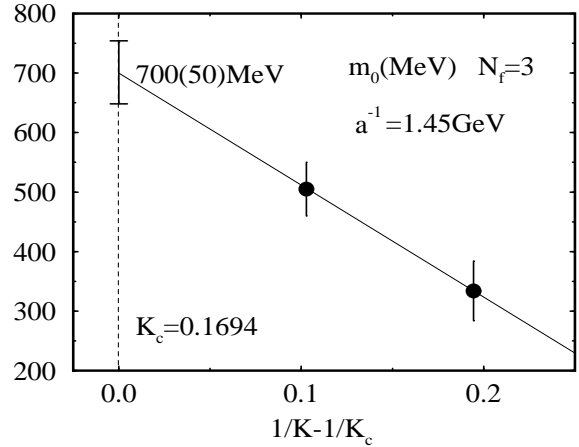}
\vskip -1.0cm }
\caption{$m_0$ in physical units as a function of $1/K-1/K_c$.}
\label{fig:mzero}
\vspace{-5mm}
\end{figure}

In Fig.~\ref{fig:etaratio} we plot the result for the ratio given in
(\ref{eq:etapropagator}) for Wilson quark action obtained with 160
configurations at $K=0.1665$.  We observe a quite clear signal for a
linear increase in $t$.  We then extract the $m_0$ parameter by a fit of form
(\ref{eq:etapropagator}) over the range $4\leq t \leq 8$.  The results at two
values of $K$ together with a linear extrapolation in $1/K$ to $K_c=0.1694$
are
shown in Fig.~\ref{fig:mzero}  where conversion to physical units is made with
$a^{-1}=1.45(2)$ GeV determined from the $\rho$ meson mass.
We refer to earlier attempts\cite{fku,IIY} that calculated the two quark loop
amplitude.  In particular the authors of Ref.~\cite{IIY}  estimated
$m_0=530$MeV
at $m_\pi/m_\rho=0.71$ and 330MeV at  $m_\pi/m_\rho=0.34$ with
$a^{-1}=1.81$GeV
from a calculation with the $\eta'$ source and sink fixed
at the spatial origin  using 10 configurations generated with a non-standard
gauge action on an $8^3\times 16$ lattice.

Our value $m_0=700(50)$MeV at $K_c$ may be compared to the \lq
experimental\rq\
value of 850MeV  deduced from the Witten-Veneziano formula\cite{Witten}.
Another comparison, more directly based on lattice QCD, is to invoke the
original $U(1)$ Ward identity relation $m_0^2=6\chi/f_{\pi}^2$ with $\chi$ the
topological susceptibility.  Using $\chi=4.28(33)\times 10^{-4}$ and
$f_\pi=0.0616(40)$ at $K_c$ we obtained on the same set of configurations as
for
$m_0$, we find
$m_0=1190(90)$MeV.

Let us now consider the question of infrared
singular terms induced by $\eta'$ loops in quenched QCD.
With our result for $m_0$ and $f_\pi$,  we find that the coefficient
of the logarithmic correction for the pion mass in (\ref{eq:chirallog})
equals 0.076 at $m_\pi/m_\rho=0.61 (K=0.1665)$.  Quenched
pion mass data presently restricted to  $m_\pi/m_\rho\simm{>}0.5$ do not show
evidence of  the logarithm with such a small coefficient\cite{SU}.   The
magnitude of  (\ref{eq:pipihairpin}) contributing to the $\pi$-$\pi$
scattering
length is even smaller.  For our Kogut-Susskind simulation, taking the Wilson
result $m_0a=0.48$ at $K_c$ as an indicative value and using measured  $m_\pi$
and $f_\pi$ we find  $\delta E=3.4\times10^{-5}$ at $m_\pi/m_\rho=0.32$, which
may be compared with the value $\delta E=1.2(4.0)\times10^{-4}$ extracted from
the slope of the double annihilation diagram.  We conclude that the possible
failure of the quenched approximation due to $\eta'$ loops does not become
manifest unless the simulation is made with a quark mass much smaller than the
value being taken in the current study.


\section{$\pi$-$N$ scattering lengths}

\begin{figure}[t]
\input psfig
\centering{
\hskip -0.0cm
\psfig{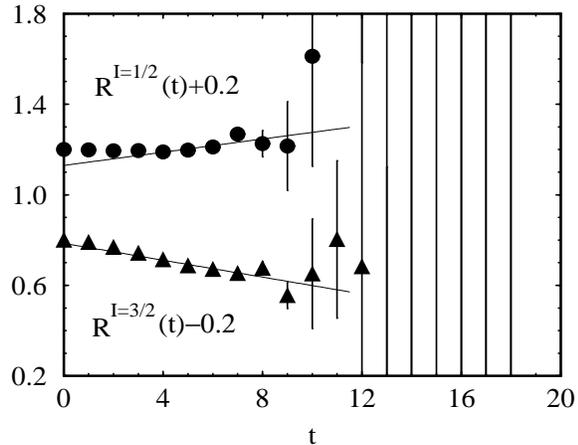}
\vskip -1.0cm  }
\caption{$\pi$-$N$ amplitudes for $I=1/2$ and 3/2.  Data are  shifted by $\pm
0.2$ to avoid overlap.}  \label{fig:piN}
\end{figure}

The wall source technique without gauge fixing does not yield good signals for
the nucleon.  We therefore used the Coulomb gauge fixing at the $t=0$ time
slice for the nucleon source to calculate $\pi$-$N$ 4-point functions.  In
Fig.~\ref{fig:piN} we show our preliminary result for the ratio $R^{I}(t)$ of
$\pi$-$N$ amplitude divided by the propagators of $\pi$ and $N$ obtained with
30
configurations with the Wilson quark action at $K=0.164$.  We observe a good
signal for $I=3/2$.  The data is much worse for $I=1/2$, marginally indicative
of a positive slope ({\it i.e.,} attraction) expected from the experiment.
Extracting the energy shift by a linear fit in $t$ over the interval $4\leq
t\leq 8$, we find for the scattering lengths
\begin{eqnarray}
a_0^{I=3/2}\cdot\frac{8\pi f_\pi^2}{\mu_{\pi N}}&=&-0.95(13)
\qquad [\;\;-1\;\;]\\
a_0^{I=1/2}\cdot\frac{8\pi f_\pi^2}{\mu_{\pi N}}&=&+1.6(0.7)
\qquad \:[\;\;+2\;\;]
\end{eqnarray}
where $\mu_{\pi N}=m_\pi m_N/(m_\pi+m_N)$ and numbers in square
brackets are predication of current algebra.     Despite a  heavy quark mass
($m_\pi/m_\rho=0.73$ at $K=0.164$) our results are consistent with current
algebra both for $I=1/2$ and $I=3/2$ channels.  The error is  large for
$I=1/2$,
however.

Clearly reducing the error is much desired for $I=1/2$.  It is also
interesting to repeat the calculation for smaller quark masses.  However, both
of these extensions are not easy in practice.   A simple
argument suggests that the errors of $R(t)$  grow exponentially
$\delta R(t)\propto e^{\alpha t}$ with $\alpha=m_N-3/2m_\pi$ or
$m_N-1/2m_\pi$ depending on the topology of quark lines\cite{Lepage}. Our
numerical data indeed exhibits such dependence with the
expected slope.  This means that a reduction of errors at large values of $t$
requires a substantial increase of
statistics.  The situation appears worse for lighter quarks since  $m_N$
remains
non-zero while $m_\pi$ becomes smaller, leading to a larger value for
the slope $\alpha$.  The only way to reduce errors would be to increase
$\beta$ with a simultaneous enlargement of the lattice size. In this regard
the
$\pi$-$N$ case turns out to be more difficult in nature than the $\pi$-$\pi$
case; for the latter the errors grow only modestly as $\delta R(t)\propto
e^{2m_\pi t}$ even for the worst case of the double annihilation
diagram\cite{pipi} whose slope becomes smaller toward the chiral limit.


\section{$N$-$N$ scattering lengths}

The $N$-$N$ scattering lengths are experimentally known to be very
large\cite{Miller};
\begin{eqnarray}
a_0({}^1S_0)&=&+20.1(4)\;{\rm fm},\\
a_0({}^3S_1)&=&-5.432(5)\;{\rm fm}.
\label{eq:nnexp}
\end{eqnarray}
As opposed to the $\pi$-$\pi$ and $\pi$-$N$ cases, the $N$-$N$ amplitudes are
not constrained by chiral symmetry.  The large values above are, therefore, a
purely dynamical phenomenon whose successful derivation is a challenge posed
to
lattice QCD.

We note that the negative sign in the  ${}^3S_1$
channel is due to the presence of the deuteron bound state (Levinson's
theorem).  This brings in an additional complication in an attempt at a
realistic
calculation of the scattering lengths, since the energy of the
lowest scattering state orthogonal to the bound-state deuteron has to
be computed to apply L\"uscher's relation to give the correct scattering
lengths.

Here we report on a less ambitious study employing quite a heavy Wilson quark
with  $m_\pi/m_\rho=0.86 (K=0.16)$.  For such a heavy quark we may expect the
deuteron to become unbound since the range of pion exchange is reduced.  In
this case the scattering lengths for both ${}^3S_1$ and ${}^1S_0$ channels
can be extracted from the lowest $N$-$N$ energy, and
we expect both to be positive in sign and large.

\begin{figure}[t]
\input psfig
\centering{
\hskip -0.0cm
\psfig{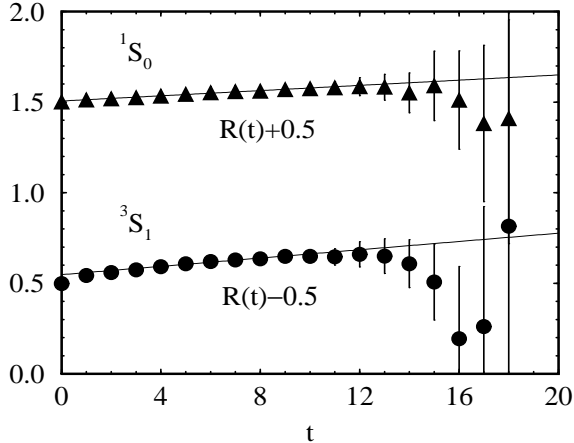}
\vskip -1.0cm }
\caption{$N$-$N$ amplitudes for ${}^3S_1$ and ${}^1S_0$ channels.  Data are
shifted by $\pm 0.5$ to avoid overlap.}  \label{fig:NN}
\vspace{-3mm}
\end{figure}

Anticipating a large scattering length we have made simulations on a $20^4$
lattice with Wilson quark action using 30 gauge configurations with Coulomb
gauge
fixing.  The result for the ratio $R(t)$ is shown in
Fig.~\ref{fig:NN}.  The positive slope corresponding to a negative energy
shift
for both channels is consistent with our expectation above.  Numerically, we
find
\begin{eqnarray}
a_0({}^1S_0)&=&1.0(3)\;{\rm fm},\\
a_0({}^3S_1)&=&1.2(2)\;{\rm fm}.
\end{eqnarray}
for the scattering length in physical units using $a^{-1}=1.45$ GeV.
Although these values are much smaller than the experimental values
they are about a factor 3 $-$ 4 larger than $\pi$-$\pi$ and
$\pi$-$N$ scattering lengths at similar quark masses,
{\it e.g.,} $a^{I=0}(\pi$-$\pi)=0.421(23)$ fm and
$a^{I=1/2}(\pi$-$N)=0.30(14)$
fm  at $K=0.164$.

We consider that the present result is encouraging; the most interesting
problem left with us is to examine whether these scattering lengths increase
rapidly as the quarks mass is reduced.
\vspace{13mm}


\end{document}